\documentclass[aps,pra,twocolumn,groupedaddress,showpacs]{revtex4}

\usepackage[latin1]{inputenc}
\usepackage[T1]{fontenc}
\usepackage{amsmath,amssymb}
\usepackage{bbm}
\usepackage{graphicx}
\usepackage{subfigure}
\usepackage{sistyle}
\usepackage{ifpdf}
\usepackage{xcolor}
\usepackage{physics}

\begin{document}

\title{Quantum estimation of a time dependent perturbation}

\author{Claus Normann Madsen}
\affiliation{Center for Complex Quantum Systems, Department of Physics and Astronomy, University of Aarhus, DK-8000 Aarhus C, Denmark.}

\author{Lia Valdetaro}
\affiliation{ Department of Physics and Astronomy, University of Aarhus, DK-8000 Aarhus C, Denmark.}

\author{Klaus M\o lmer}
\affiliation{Center for Complex Quantum Systems, Department of Physics and Astronomy, University of Aarhus, DK-8000 Aarhus C, Denmark.}
\affiliation{Aarhus Institute of Advanced Studies, University of Aarhus, DK-8000 Aarhus C, Denmark.}

\begin{abstract}

We analyze the estimation of a time dependent perturbation acting on a continuously monitored quantum system. We describe the temporal fluctuations of the perturbation by a Hidden Markov Model, and we combine quantum measurement theory and classical filter theory into a time evolving hybrid quantum and classical trajectory. The  forward-backward analysis that permits smoothed estimates of classical Hidden Markov Models has a counterpart in the theory of retrodiction and Past Quantum States. As a specific example, we apply our hybrid trajectory and Past Quantum State theory to the sensing of a fluctuating magnetic field by microwave interrogation of a single quantum spin.

\end{abstract}

\maketitle

\section{Introduction}

Quantum systems, such as atoms, are ideal time keepers and are sensitive over broad bandwidths to perturbations such as magnetic and electric fields. Unique quantum features e.g. coherent superposition states, squeezing, entanglement and quantum phase transitions are being employed to maximize measurement sensitivity \cite{Maccone2,Maccone}. The use of quantum systems for near field and precision measurements is thus a quantum technology with near term prospects to benefit society. Quantum measurement outcomes are governed by Born's rule and hence by a fundamental randomness that plays important roles for the precision and sensitivity of any quantum metrological protocol. In the case of continuous measurements over time, the random outcomes are accompanied by measurement back action which quenches the quantum system and imposes nontrivial correlations in the outcome of subsequent measurements. This may yield higher sensitivity to perturbations than one would infer from their influence on the expectation values of system observables \cite{Mabuchi,Gambetta,Gammelmark,Kiilerich}.

In this article we employ quantum trajectory theory \cite{Belavkin,Carmichael}, i.e., stochastic master equations that describe the evolution of a system density matrix conditioned on both unitary and dissipative evolution and on the random outcomes of measurements on the system. We incorporate an unknown classical perturbation by a Hidden Markov Model (HMM) \cite{HMMreview,HMMMabuchi,HMMGammelmark}  with hidden (classical) states $n$. The system jumps between states with the rates $r_{n\rightarrow n'}$, and the state dynamics dictate the dynamics of the strength of the perturbation on the quantum system, e.g., through the value of terms, $\Delta(t)=\Delta_{n(t)}$, in the system Hamiltonian. Measurements on the system probe its evolution and may hereby reveal the properties of the perturbations acting on the system.

The so-called forward-backward algorithm makes use of a full measurement record to determine the time dependent state of a classical HMM. In \cite{HMMMabuchi,HMMGammelmark}, this approach was shown to also apply to the analysis of data from incoherent quantum systems, and in \cite{TsangPRL,TsangPRA,Cheng} probing of time dependent perturbations with Gaussian noise correlations by quantum systems restricted to Gaussian states was shown to closely follow the classical theory of Mayne-Fraser-Potter two-filter smoothers \cite{Kalman}. Here, we discuss a more general class of perturbations and quantum probes that require a full density matrix formalism. The effect of measurements is incorporated by a stochastic master equation, and the retrodictive power offered by later sequences of measurement records is captured by forward and backward quantum filter equations in the so-called Past Quantum State theory  \cite{PQS}.  

The HMM may represent a variety of fluctuating parameters. Here we present the example of a fluctuating magnetic field which causes fluctuations in the transition frequency of a resonantly driven two-level spin system in a microwave cavity. Such systems are subject to intense research which aim to detect and ultimately control the dynamics of single spin impurities in architectures compatible with superconducting quantum computer architectures \cite{HaikkaBertet,Probst}. Following \cite{HaikkaBertet}, we assume the cavity is probed by a coherent microwave field, and we simulate the noisy outcome of homodyne detection of the  reflected signal to estimate the unknown time dependent magnetic field. 


 \begin{figure}[h!]
    \centering
    \includegraphics[width=0.3\textwidth]{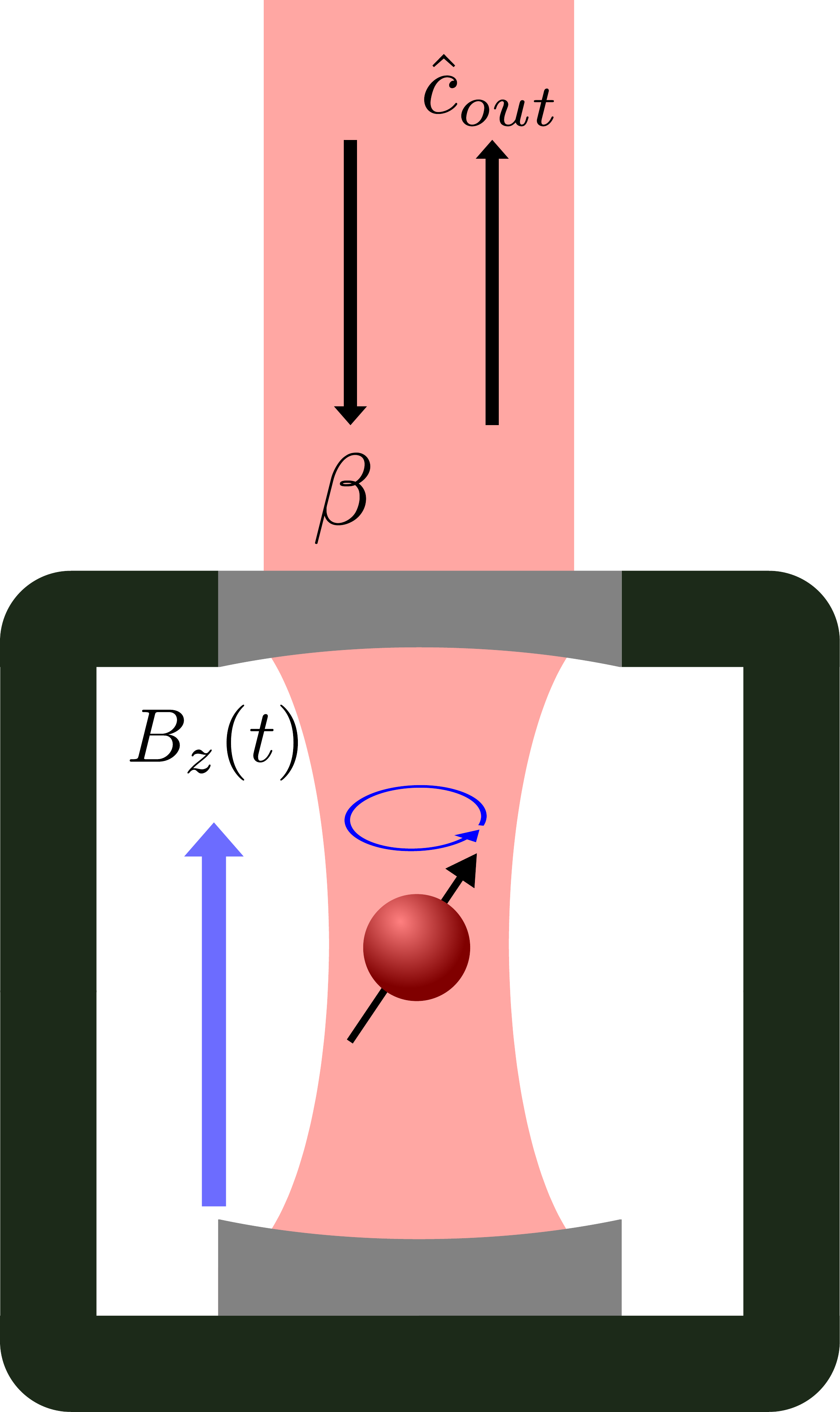}
    \caption{Schematic set-up for detection of a time varying magnetic field $B_z(t)$. A single spin, e.g., a dopant spin-1/2 in a solid state material, interacts near resonantly with a microwave cavity field. When a magnetic field shifts the energy difference between the spin eigenstates, it changes the  amplitude and phase of the reflected field from the cavity. See \cite{HaikkaBertet} for an implementation geometry that maximizes the spin-cavity coupling.} 
    \label{fig:model}
\end{figure}

The article is organized as follows. In Sec. II, we present our physical sensing device composed of a two-level spin system in a cavity which is probed by a classical field \cite{HaikkaBertet}. We eliminate the cavity field and present the stochastic master equation for the two-level system subject to continuous homodyne detection of the cavity output. In Sec. III, we introduce the joint stochastic master equation describing the conditional classical and quantum dynamics of the HMM and the probe quantum system, and we provide the explicit theory for the case of a fluctuating magnetic field and a two level spin system. In Sec. IV, we recall the past quantum state formalism \cite{PQS} and we discuss how it leads to a time dependent probability distribution of the HMM variable governing the perturbation on the quantum system. We provide simulations that quantify the precision of the estimation of the perturbing field and we demonstrate the advantage of using the full measurement record to determine the perturbation at any time. Sec. V concludes and presents an outlook. 

\section{Description of the model}

\subsection{A resonantly driven spin system}

Following \cite{HaikkaBertet}, we consider the situation of a two-level system in a cavity subject to a classical driving field at the frequency $\omega_d$, $\beta_{in} = \beta e^{-i\omega_d t}$, see Fig.~\ref{fig:model}. The reflected signal is subject to homodyne detection and will thus reveal the interaction of the two-level system with a time dependent magnetic field that shifts its resonance frequency with respect to the one of the cavity and the classical drive field.   
The cavity is damped faster than the other dynamical time scales, and the cavity field degrees of freedom can therefore be adiabatically eliminated. In a frame rotating with the driving field, the Hamiltonian of the qubit and cavity can be written as
\begin{equation}
H = \Delta_r \hat a^\dag \hat a + i \sqrt{2\kappa_1} (\beta \hat a^\dag - \beta ^* \hat a) + \frac{\Delta_s}{2} \hat \sigma_z + g(\hat \sigma_+ \hat a + \hat \sigma_- \hat a^\dag), \label{initial_Hamiltonian}
\end{equation}
where $\Delta_r$ is the detuning between the cavity and the driving field, $\hat a$ ($\hat a ^\dag$) is the annihilation (creation) operator of the photonic field in the cavity,
$\kappa_1$ is the amplitude decay rate of the cavity field due to transmission to the outgoing field, and $\beta$ is the amplitude of the driving field. The atom or spin system is characterized by the $\hat \sigma_z$  and $\hat \sigma_+$ ($\hat \sigma_-$) Pauli-$Z$ and raising (lowering) operators, the detuning $\Delta_s$ between the qubit spin and the driving field, and the coupling strength $g$ with the quantized cavity field.

The system is subject to damping processes represented by Lindblad operators 
\begin{align}
\hat c_1 =& \sqrt{2\kappa} \hat a , \\
\hat c_2 =& \sqrt{\gamma_{dec}} \hat \sigma_- \label{c2} ,\\
\hat c_3 =& \sqrt{\frac{\gamma_\phi}{2}} \hat \sigma_z. \label{c3}
\end{align}
Here, $\hat c_1$ represents the decay of the photon field in the cavity either by transmission (rate $\kappa_1$) or losses (rate $\kappa_L$, $\kappa=\kappa_1+\kappa_L$). $\hat c_2$ denotes the possible decay of the qubit excitation outside of the cavity mode, and $\hat c_3$ denotes dephasing of the qubit.

The outgoing field from the cavity is given by input-output theory as
\begin{equation}
\hat c_{out} = \sqrt{2\kappa_1}\hat a - \beta.
\end{equation}
The homodyne detection of the outgoing  field causes measurement back action and is represented by a stochastic term in the master equation involving the operator $\hat c_{out}$. 

\subsubsection{Adiabatic elimination of the bad cavity}

The cavity facilitates efficient probing of the spin system, but in the bad cavity limit $g \ll \kappa$, the cavity degrees of freedom follow the evolution of the spin closely and may thus be adiabatically eliminated from the formalism.
This considerably simplifies the system of equations.

The cavity field operator $\hat a$ obeys the Heisenberg equation of motion,
\begin{equation}
\pdv{\hat a}{t} = - i \Delta_r \hat a + \sqrt{2\kappa_1}\beta - ig \hat \sigma_- - \kappa \hat a + \hat F,
\end{equation}
where $\hat F$ is the input vacuum field Langevin noise with vanishing expectation value. Omitting the noise and setting $\pdv{\hat a}{t}=0$ permits elimination of $\hat a$ under the assumption of adiabatic following of the qubit in the frame rotating with frequency $\omega_s$,
\begin{equation}
\hat a = \frac{\sqrt{2\kappa_1}\beta}{\kappa + i \Delta_r} - \frac{i g \hat \sigma_-}{\kappa + i (\Delta_r - \Delta_s)}.
\end{equation}
Inserting this expression in \eqref{initial_Hamiltonian} and \eqref{c3} we obtain the effective qubit Hamiltonian
\begin{equation} 
H = \frac{\Delta_s}{2} \hat \sigma_z + g(\alpha \hat \sigma_+ + \alpha^* \hat \sigma_-) - \epsilon_s \hat \sigma_+ \hat \sigma_-, \label{Hadel}
\end{equation}
where $\alpha=\frac{\sqrt{2\kappa_1}\beta}{\kappa + i \Delta_r}$ and $\epsilon_s = \frac{g^2 (\Delta_r - \Delta_s)}{\kappa^2 + (\Delta_r - \Delta_s)^2}$, and 
\begin{equation}
\hat c_1 = \sqrt{\gamma_p}\hat \sigma_-, \label{c1}
\end{equation}
with the Purcell decay rate
\begin{equation}
\gamma_p = \frac{2 g^2 \kappa}{\kappa^2 + \left(\Delta_r - \Delta_s\right)^2}.
\end{equation}

Finally, the out-coupled field is conveniently represented by the operator
\begin{equation}
\hat c_{out} =  \sqrt{2\kappa_1}\left( \alpha - \frac{i g \hat \sigma_-}{\kappa + i (\Delta_r - \Delta_s)} \right) - \beta .
\label{cout}
\end{equation}

\subsubsection{Homodyne measurement master equation}
A quantum system subject to homodyne detection evolves according to the stochastic master equation (SME) \cite{SME1, SME2}, $\tilde\rho(t+dt)= \tilde\rho(t) + d\tilde\rho$ where 
\begin{equation}
d\tilde \rho = \mathcal{L} \tilde \rho dt + \mathcal{X}_\Phi \tilde \rho dY_t. \label{forward_homodyne_master_equation}
\end{equation}
The tilde indicates that the density matrix is not normalized. The first term yields the usual Lindblad master equation terms
\begin{equation} \label{LME}
\mathcal{L}\rho = -i\comm{\hat H}{\rho} + \sum_i\mathcal{D}[\hat c_i]\rho,
\end{equation}
where $H$ and $\hat c_i$ are the Hamiltonian and Lindblad damping operators presented in the previous subsection with $\mathcal{D}[\hat A] \rho = 	\hat A \rho \hat A^\dag - \frac{1}{2} (\hat A^\dag \hat A\rho + \rho\hat A^\dag \hat A)$. The second term in \eqref{forward_homodyne_master_equation} represents the back action due to the homodyne measurement of the output field, with
\begin{equation}
\mathcal{X}_\Phi \rho = \sqrt{\eta} \left(\hat c_{out} e^{-i \Phi} \rho + \rho \hat c_{out}^\dag e^{i \Phi} \right),
\end{equation}
where $\Phi$ is the phase of the local oscillator used for homodyne detection, and $\eta$ is the measurement efficiency. $dY_t$ is the measured homodyne signal in the time-step $dt$ and is composed of a mean and a fluctuating contribution,
\begin{equation}
dY_t = \Tr[ \mathcal{X}_\Phi \rho(t) ] dt + dW_t, \label{eq:dYt}
\end{equation}
where $dW_t$ is a Wiener increment of zero mean and variance $dt$. The mean value is governed by the trace of $\mathcal{X}_\Phi \rho(t)$, evaluated with the (normalized) density matrix $\rho(t)$.

\section{Hybrid classical and quantum description of the unknown perturbation and the quantum probe}
  
\subsection{Hidden Markov model of the perturbation}

Hidden Markov Models (HMMs) are powerful tools to analyze time series of data ranging from the natural sciences to linguistics and sociology \cite{HMMreview}. In classical applications, the model describes a hidden parameter that explores a discrete space of values (states) $n$ by randomly jumping with rates $r_{n\rightarrow n'}$ among them. The assumption is that the hidden parameter governs the evolution of the system of interest and that observation of some of its degrees of freedom gives rise to a detection signal with statistics that depend on the state occupied by the hidden Markov parameter. The change of character over time of the observed signal can thus be ascribed to changes among the value for $n$, and if the probability of a given signal outcome is known for each state, Bayes' rule, together with the rate equations, permits evaluation of the probability $P(n)$ that the system occupies the different states conditioned on the measurement. Such models show a rich variety of behaviors and both the signal probabilities, the transition rates -  and even the number of states - can be treated as fitting parameters to develop efficient models for various kinds of signals \cite{HMMMabuchi,HMMGammelmark}.

We have previously used HMMs to describe incoherent dynamics in quantum physics, such as off-resonant excitation and spontaneous decay causing transfer between hyperfine atomic states monitored by cavity transmission signals \cite{HMMGammelmark}, and fluctuations of the photon number in a microwave cavity subject to dispersive interactions with a sequence of probe atoms \cite{Dotsenko}. In this article our aim is to employ the hidden Markov model to represent an unknown time dependent physical perturbation. 

In our example, the discrete variable $n$ represents the fluctuating candidate value $B_n$ of a magnetic field, which causes energy shifts of spin Zeeeman sub-levels and hence a variation of the detuning $\Delta_s$ in the model described in the previous section. For simplicity, we merely include this modification of $\Delta_s$ in the expressions (8,10,11), obtained after the adiabatic elimination,.

\subsection{Extending the density matrix to include the fluctuating perturbation}

We can embed the unknown value of the fluctuating classical field in an effective  quantum state description that  makes use of the fact that the HMM is equivalent to a quantum system restricted to incoherent jumps between the states $n$. In such a description, transitions between the (classical) states can be represented by quantum jump Lindblad terms
\begin{equation}
\hat J_{nn'} = \sqrt{r_{nn'}} \dyad{n'}{n} , \label{Linblad_B_Jump_Operator}
\end{equation}
and we combine the HMM and the two-component spin density matrix by the tensor product
\begin{align} \label{block}
\tilde{\rho} &= \sum_n \tilde{\rho}_n \otimes \dyad{n}{n}\nonumber \\
&= \begin{bmatrix}
\tilde{\rho}_1 & & & 0\\
& \tilde{\rho}_2 & & \\
& & \ddots & \\
0 & & & \tilde{\rho}_N
	\end{bmatrix}.
\end{align}
Each $\tilde{\rho}_n$ term is an unnormalized $2\times 2$ density matrix on a two-level system Hilbert space, correlated with the diagonal elements $\dyad{n}{n}$, i.e., the unknown classical value $B_n$ of the magnetic field. We emphasize that the unknown perturbation is now treated in complete equivalence with an additional quantum degree of freedom of an enlarged system. The full density matrix of the enlarged system solves the stochastic master equation \eqref{LME}, where in addition to the terms acting on each $\tilde{\rho}_n$ component (note that $c_i \rightarrow c_i \otimes \sum_n \dyad{n}{n}$, and the Hamiltonian $H= \sum_n H_n\otimes \dyad{n}{n}$ depends on $n$ through the detuning parameter in \eqref{Hadel}), Lindblad-like terms, $\sum_{nn'}\mathcal{D}[I\otimes\hat J_{nn'}]\rho$ distribute the atomic system density matrix elements among the sub-blocks in \eqref{block} to represent the probabilistic evolution of the hidden Markov parameter (the magnetic field).

Normalizing $\tilde{\rho}$ to unit trace is equivalent to the scaling of all $\rho_{n}=\tilde{\rho}_n/\Tr(\tilde{\rho})$, which yields the probability distribution of the magnetic field as the population of the corresponding hidden state $n$, $P(B_n)=\Tr(\rho_n)$. 

The dynamics of the full system density matrix \eqref{block}, subject to the various elements of the master equation, and in particular to the measurement back action, is a direct implementation of the Belavkin filter idea \cite{Belavkin}. The redistribution of probability among the candidate values $B_n$ is equivalent to a Bayesian update, since the change in norm of each $\tilde{\rho}_n$ due to the term in \eqref{forward_homodyne_master_equation} proportional to the stochastic measurement outcome $dY_t$ directly represents the probability of that outcome conditioned on the state. The trace of the $\tilde{\rho}_n$ components that are more compatible thus experience a relative increase with respect to the ones that are less compatible with the outcome measurement. 

\subsection{Specification of system for sensing of magnetic field}

Subjecting the system to a magnetic field $B_z(t)$ changes the qubit energy difference through the Zeeman effect,
\begin{equation}
H_{Z} = -\frac{g_{qB}\mu}{2}B_z(t) \hat \sigma_z,
\end{equation}
where $g_{q_B}$ is the Lande factor and $\mu$ is the electron magnetic moment.  

For a range of discrete values $B_n$ of the magnetic field, the system Hamiltonian thus attains the values
\begin{equation}
\hat H_{n} = \frac{\Delta_s(B_n)}{2}\hat \sigma_z + g \left(\alpha \hat \sigma_+ + \alpha^* \hat \sigma_-\right) - \epsilon_s(B_n) \hat \sigma_+ \hat \sigma_-,
\end{equation}
where
\begin{equation}
\Delta_s(B_n) = \Delta_s - g_{q_B}\mu B_n.
\end{equation}
We recall that also the stochastic measurement term and the Lindblad damping terms depend on the value of $B_n$ in our calculations.

While our treatment is general, we shall study the example of a hidden Markov model known as the Ehrenfest dog-flea model \cite{hauert2004dogs}, describing the evolution of the difference in the number of randomly jumping fleas between two dogs. This model corresponds to a discrete version of the stochastic Ornstein-Uhlenbeck process, a diffusion process with an approximate Gaussian steady state distribution.

\subsection{Numerical example}

The simulations were done using the \verb!C++! library  Armadillo \cite{armadillo, armadillo_sparse}. In the simulations, we use natural units of $\hbar = 1$ and we assume  the spin decay and dephasing rates $\gamma_{dec}=\gamma_\Phi \equiv \gamma$, and a single-photon coupling strength, $g = 2 \gamma$ and cavity field decay rates $\kappa = \kappa_1 = 10 \gamma$. The driving field and the cavity are assumed to be resonant, $\Delta_r = 0$, with the strength $\beta = \sqrt{\gamma}$, and the candidate B-field states yield a range of spin detunings, $\Delta_n = -\frac{g_{q_B}\mu}{2}B_n \in [-2,..,2] \gamma$. Our choice of parameters leads to a Purcell decay rate with a weak dependence on the detuning  (and thus on the magnetic field)  between $\gamma_p(\Delta_n = 0) = \frac{4}{5} \gamma$ for the central value of the magnetic field and $\gamma_p(\Delta_n = \pm 2\gamma) = \frac{10}{13} \gamma$ at the extremum  values considered in our  estimate.  For the chosen parameters, the adiabatic elimination of the cavity state-space is valid. 

The fluctuating magnetic field  is modelled by the dog-flea model of $N$ fleas jumping at a rate $p$ between two dogs. For large $N$ values, the number of fleas on one of the  dogs, $n$, can be approximated by a continuous parameter $x$, subjected to the Ornstein-Uhlenbeck process with restoring and white noise terms, $dx = -2p (x-N/2) dt + \sqrt{pN} dW$. In our model study, we assume $p = 0.02\gamma$ for the characteristic rate of fluctuations among $N+1=25$ different values of the magnetic field and, hence, of the detuning $\Delta_n$.

We have simulated the time dependent variation  of $n(t)$, which is used to update the Hamiltonian $H_{n(t)}$ and measurement and damping terms that drive the quantum system. A typical homodyne detection record can subsequently be constructed by simulation of the $dW$ white noise in the expression for $dY_t$ \eqref{eq:dYt}, which is inserted in the stochastic master equation \eqref{LME} for the small system. Finally, the simulated homodyne signal can be applied in the master equation for the extended density matrix \eqref{block}, where the parameter $n$ is treated as an unknown,  and we can extract the  probability distribution $P(n,t)=\Tr(\rho_n(t))$, inferred  from the renormalized density matrix. 

\begin{figure}[h!]
    \centering
    \includegraphics[width=0.4\textwidth]{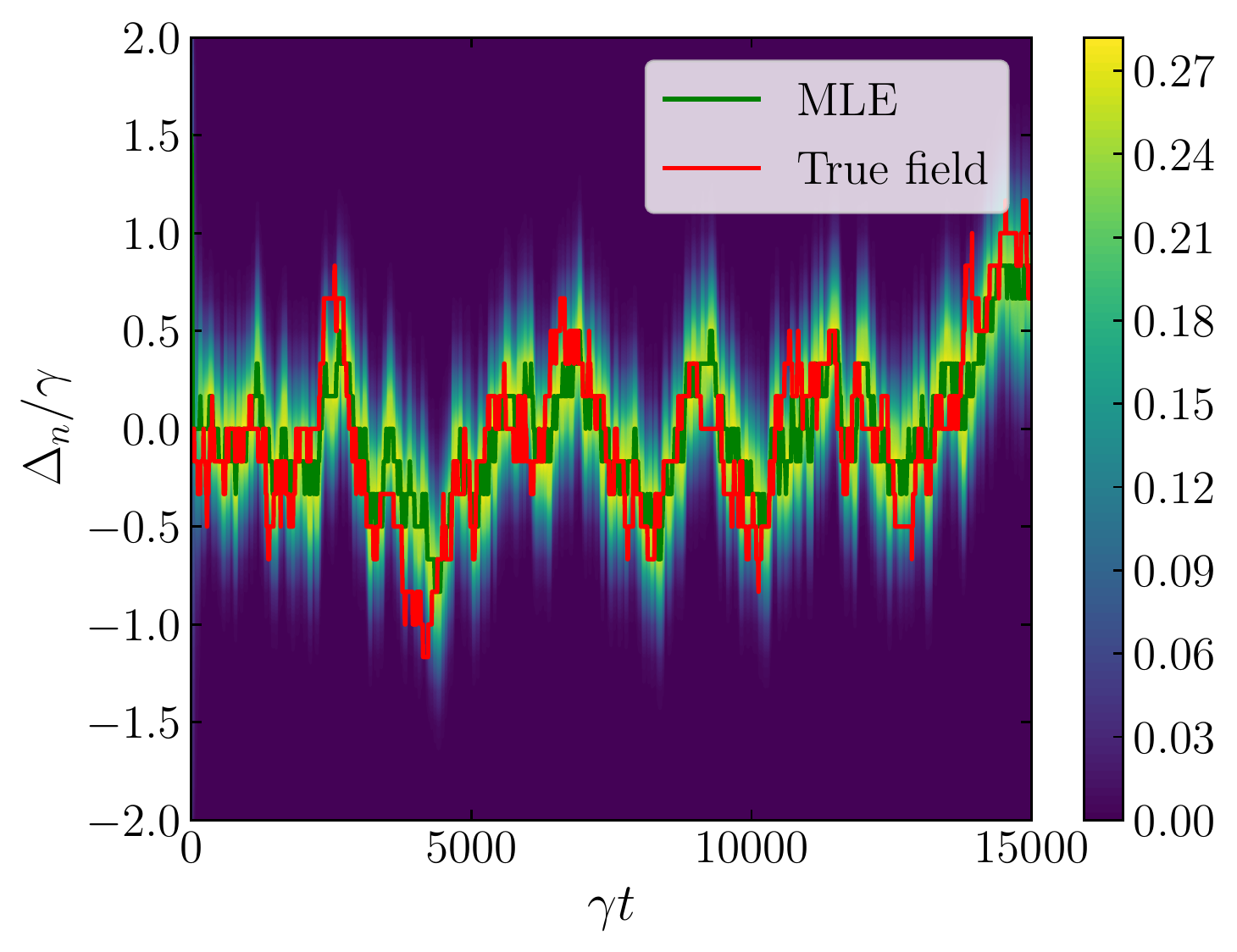}
    \caption{Conditional probability distribution for the time dependent detuning shifts experienced by a spin qubit system subject to a fluctuating magnetic field. The red curve shows the true simulated variation, while the bright area and the central green curve show the  distribution and most likely value of the field inferred by probing of the spin dynamics. The root mean square deviation of the maximum likelihood estimate from the true value of the detuning is 0.26$\gamma$. The parameters are given in the text.}
    \label{fig:filtering}
\end{figure}

In Fig.~\ref{fig:filtering}, the red curve shows the classically simulated realization of the time dependent detuning. The bright colored area indicates the width of the inferred probability distribution, and the green curve shows the discrete value $\Delta_n$ corresponding to the maximum value of $P(n,t)$. The estimated value of the detuning follows the exact value in an overall sense, but it shows fluctuations that are uncorrelated with the fluctuations in the true value. The homodyne detection is a noisy process and it requires integration over time to acquire a significant signal to noise ratio. A closer look hence shows that variations in the estimated value generally lag behind the variations in the true value of the external perturbation. 

With the specified parameters, the unobserved dog-flea model yields a distribution of detunings (around zero) with a standard deviation $0.41\gamma$. The difference between the simulated value and the maximum likelihood estimator fluctuates with a time-averaged root mean square value of $0.26 \gamma$. As expected, this is close to the value, $0.27 \gamma$, inferred as the square root of the average variance of $P(n,t)$ (in detuning units), depicted by the bright colored area in Fig.~\ref{fig:filtering}. 

\section{Forward-backward estimate by the Past Quantum State formalism}
In the previous section we showed how the Hidden Markov Models and the Bayesian update mechanism can be embedded in an extended quantum trajectory description. As future measurement statistics are similarly governed by the state of the system, the properties of the state at time $t$ are also correlated with the measurement outcomes {\it after} $t$ \cite{Vivi}. The Markovian description of measurements on the emitted field  permits identification of recursive equations, one propagating the state $\rho$ forward in time conditioned on previous signal data, the other propagating a matrix $E$ backwards towards the time $t$ conditioned on the measurement data acquired after $t$. Combination of these two matrices represent a generalization of the so-called forward-backward or $\alpha-\beta$ analysis of HMMs \cite{HMMMabuchi,HMMGammelmark} to the quantum case \cite{PQS}.

\subsection{The Past Quantum State}

For a brief explanation, the quantum trajectory conditional dynamics can be written as a deterministic evolution intercepted by stochastic elements which can in turn be written as POVM operators acting on the state (multiplying the density matrix from the left and right by operators $\hat{M}_m$ and $\hat{M}_m^{\dagger}$ according to the measurement outcomes, or applying weighted sums of such terms). The joint probability of all data in a long measurement record is then simply given by the trace of an expression with the corresponding POVM operators acting sequentially from left and right on the density matrix. Some of these operators represent the known action of measurements until time $t$, some represent the action of a potentially unknown measurement at time $t$, and some represent the known results obtained after $t$.  The former product of terms yields the (unnormalized) density matrix $\rho(t)$, while the cyclic properties of the trace permits reordering and combination of all the latter terms in a single matrix $E(t)$. The resulting expression for the outcome probability of an arbitrary measurement {\it at time} $t$ with POVM operators $\hat{\Omega}_m$ then reads
\begin{equation} \label{pqsprob}
P(m)=\frac{\Tr(\hat{\Omega}_m \rho(t)\hat{\Omega}_m^{\dag} E(t))}
{\sum_{m'} \Tr(\hat{\Omega}_{m'} \rho(t)\hat{\Omega}_{m'}^{\dag} E(t))}, 
\end{equation}
where the denominator acts as a normalization factor. 
We see that the later measurement outcomes yield an explicit deviation from the conventional expression, $P(m)=\Tr(\hat{\Omega}_m \rho(t)\hat{\Omega}_m^{\dag})$. The matrix character of the expression and the fact that $\rho$, $\hat{\Omega}_m$ and $E$ may not have common eigenstates, leads to some quantative differences with the results for the classical HMM \cite{HMMMabuchi,HMMGammelmark}.

The effective stochastic master equation for $\rho$, subject to homodyne detection, is given in \eqref{forward_homodyne_master_equation}, and is equivalent to application of the POVM elements \cite{SME1,SME2}
\begin{equation}
\hat M_{dY_{t}} = \frac{e^{\frac{-dY_t^2}{4dt}}}{^{4}\sqrt{2\pi dt}}\left(1 -i\hat H dt - \frac{\hat c_{out}^\dag \hat c_{out}}{2}dt + \sqrt{\eta}e^{-i\Phi}\hat c_{out} dY_t\right),
\end{equation} 
associated with the homodyne signal $dY_t$. 
The evolution equation for $E$ involves the adjoint operation (from the cyclic transfer of the $\hat{M}_{dY_t}$ within the trace expression for the signal record probability)
\begin{equation}
E_{t-dt} = \hat M_{dY_{t-dt}}^\dag E_t \hat M_{dY_{t-dt}}.
\end{equation}
Combining this term with the unitary and damping terms  yields the backward, adjoint 'master equation'\cite{PQS}  $E(t-dt)=E(t) + dE(t)$, where $dE(t)$ is conditioned on the homodyne measurement signal $dY_{t-dt}$, 
\begin{align}
dE =& i\comm{\hat H}{E} + \sum_i \mathcal{D}^\dag[\hat c_i]E \nonumber\\
&+ \sqrt{\eta}\left(e^{i\Phi} \hat c_{out}^{\dag}  E + E\hat c_{out}e^{-i\Phi} \right)dY_{t-dt},
\end{align}
and $\mathcal{D}^\dag$ denotes the adjoint Lindblad term dissipator
\begin{equation} \label{ladj}
	\mathcal{D}^\dag[\hat A] E = \hat A^\dag E \hat A - \frac{1}{2} \acomm{\hat A^\dag \hat A}{E}.
\end{equation}

The incorporation of the classically fluctuating parameter by the HMM is done for $E(t)$ as for $\rho(t)$, i.e., we introduce the extended block-diagonal operator $E(t)$,
\begin{align} \label{blockE}
E =& \sum_n E(n) \otimes \dyad{n}{n}\nonumber \\
=& \begin{bmatrix}
E_1 & & & 0\\
& E_2 & & \\
& & \ddots & \\
0 & & & E_N
\end{bmatrix},
\end{align}
and its evolution incorporates the conjugate Lindblad terms cf., \eqref{ladj} with $\hat{A}=\hat{J}_{nn'}$. Unlike the forward HMM for which these transitions rates yield a steady state distribution that balances the transitions among the HMM states $n$, the backward adjoint rate equations have the uniform distribution as its steady state, as readily seen by inserting the identity matrix for $E$ in \eqref{ladj}. In the presence of probing, however, the $E$ matrix is nontrivial and contributes to the estimate of the HMM parameter.    

\subsection{PQS probabilities for the classically fluctuating HMM parameter}

Solving the equations for $\rho(t)$ and $E(t)$ with the specified Hamiltonians, Lindblad operators and stochastic increments, we are in possession of block matrices \eqref{block} and \eqref{blockE}. The probability of the $n^{th}$ state of the HMM is given by \eqref{pqsprob}, with the projective operator
$\hat{\Omega}_n = \dyad{n}{n}$, which up to normalization gives the result,    
\begin{align}
P_{PQS}(B_n, t) =& \Tr(\dyad{n}{n} \rho(t) \dyad{n}{n} E(t)) \nonumber\\
=& \Tr(\rho^{(n)}(t) E^{(n)}(t))  \nonumber \\
=& \rho^{(n)}_{gg}(t)E^{(n)}_{gg}(t) + \rho^{(n)}_{ee}(t)E^{(n)}_{ee}(t) \nonumber \\
&+ \rho^{(n)}_{ge}(t)E^{(n)}_{eg}(t) + \rho^{(n)}_{eg}(t)E^{(n)}_{ge}(t) \label{PQS-trace}.
\end{align}
Note that unlike the simple relation, $P(B_n,t)=\Tr(\rho^{n}(t))=\rho^{(n)}_{gg}(t)+ \rho^{(n)}_{ee}(t)$, the (unnormalized) retrodicted probability depends on both the populations and the coherences in the system density matrix $\rho_n(t)$ and effect matrix $E_n(t)$, and it does not simply factor into a product of terms conditioned on early and later detection outcomes. In general the PQS estimate  improves the estimate, essentially because it uses more  data (time windows both before and after $t$) for the estimate. That would suggest a factor two reduction of the variance of the estimator, but due to correlations in the signal data and the matrix character of the expression \eqref{PQS-trace}, the improvement is not given by a simple factor \cite{Cheng}.

\begin{figure}
    \centering
    \includegraphics[width=0.4\textwidth]{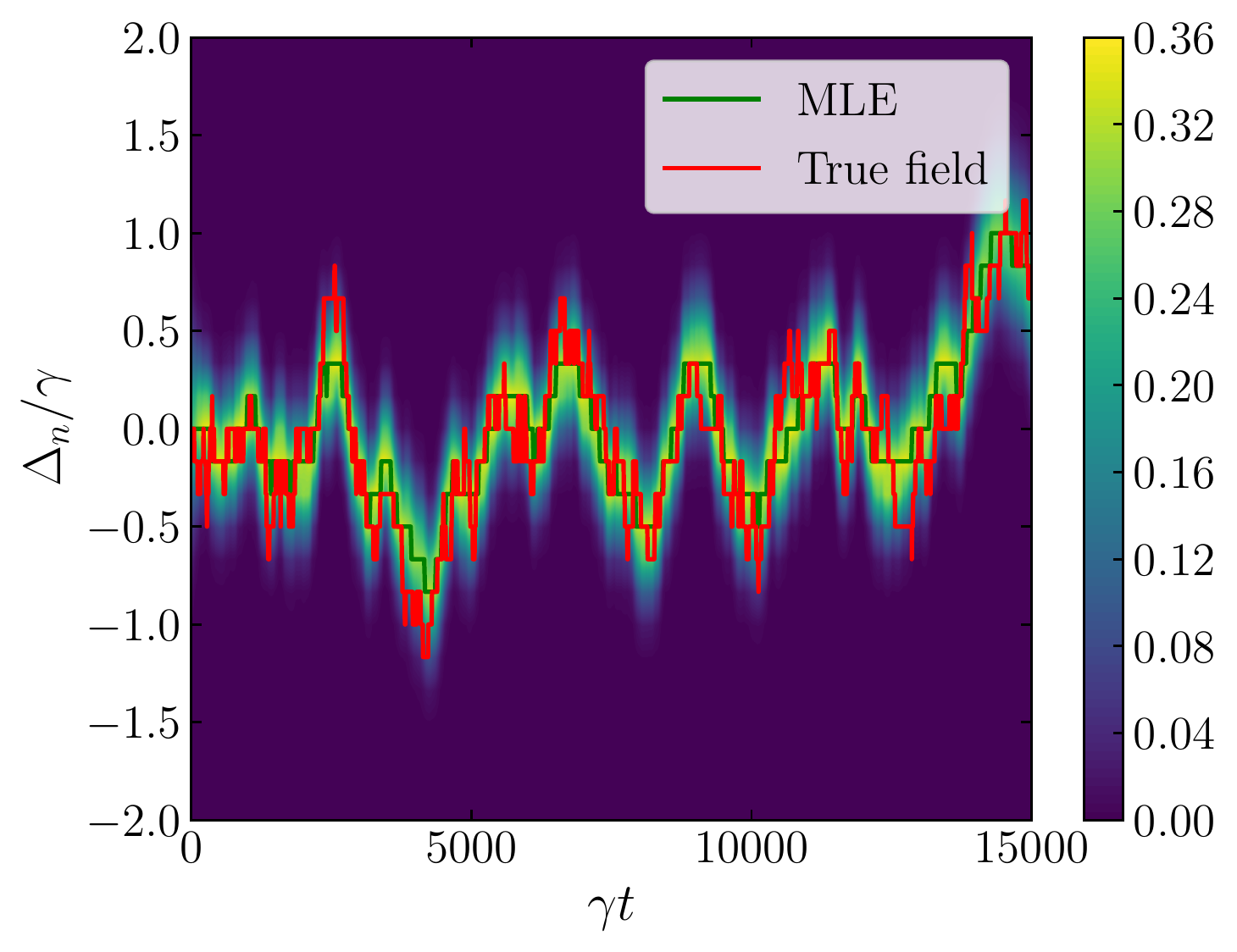}
    \caption{Past quantum state probability distribution for the detuning of the spin transition imposed by a randomly varying  magnetic field. The red curve shows the true simulated variation, while the bright area and the central green curve show the inferred distribution and most likely value. The parameters are given in the text. The root mean square deviation of the maximum likelihood value from the actual value, averaged over time, is 0.20$\gamma$ in good ageement with the standard deviation of the conditional probability distribution.}
    \label{fig:PQS}
\end{figure}

Figure \ref{fig:PQS} shows an analysis of the same simulated measurement data used to produce Figure \ref{fig:filtering}, but here we apply the Past Quantum State formalism and Eq.\eqref{PQS-trace} to provide the probability distribution and the maximum likelihood estimator for the detuning of the spin transition. The figure shows that the conditional probability of the time dependent  value of the classical pertubation is tighter, and the maximum likelihood estimator (green curve) is generally closer to the correct value. In particular, we observe no appreciable delay, as in Fig.~\ref{fig:filtering} between the true value and the estimator, but the estimator misses sudden peaks and depressions in the true signal. This can be understood as an effect of smoothing, and is formally rooted in the way that the random signal $dY_t$ at any given time contributes with similar effect to future values of $\rho$ and earlier values $E$, and does hence not change the expression \eqref{PQS-trace} appreciably as time progresses between $t-dt,\ t$ and $t+dt$ \cite{PQS}.  

\subsection{Comparison of forward and PQS estimation for different probe strengths}

\begin{figure}[h!]
    \centering
    \includegraphics[width=0.4\textwidth]{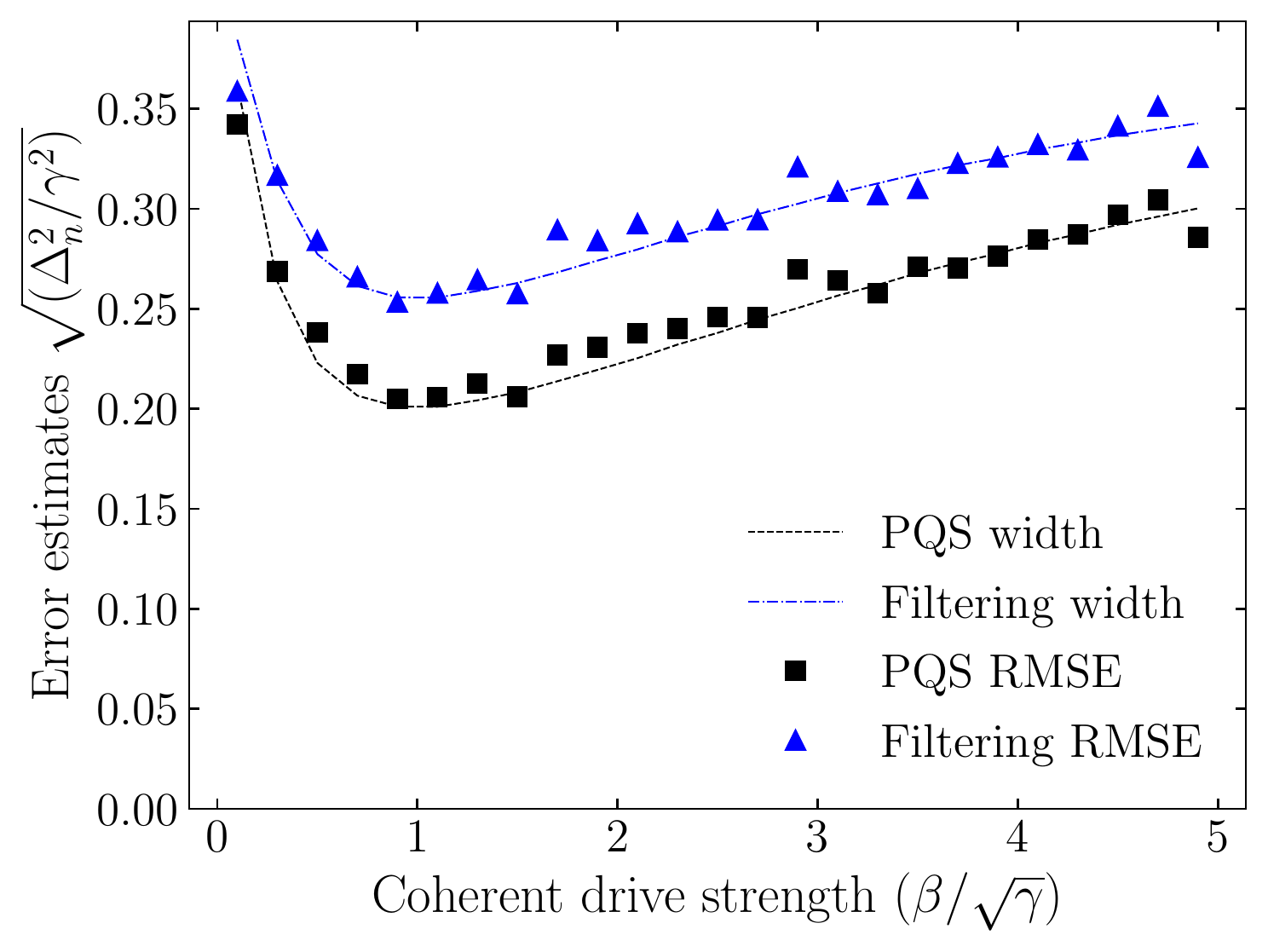}
    \caption{Error estimates defined as the square root of the time averaged variances of the forward filter and PQS distributions (upper and lower solid curves respectively)  and of the mean squared errors of the maximum likelihood estimators with respect to the true value of the fluctuating detuning. The results are shown as functions of the probe field amplitude $\beta$}
    \label{beta_figure}
\end{figure}

From the above analysis we understand how the time needed to accumulate sufficient statistics prevents instant following of changes in the fluctuating magnetic field. Both the filter analysis and the Past Quantum State analysis succeed better if these fluctuations are slow. To quantify this better, we have carried out simulations with different values of the probe strength relative to the rate parameter of the fluctuating field.
The results are summarized in Fig.~\ref{beta_figure}, which was made using individual trajectories of duration $T = 200000\gamma^{-1}$ for different values of $\beta$. In the figure the curves show the square root of the time averaged variance of the conditional distribution of $\Delta_n/\gamma$, while the symbols show the square root of the time average of the squared difference between the maximum likelihood estimate and the true, simulated value. The upper curve and symbols pertain to the (forward) filter approach and the lower curve and symbols pertain to the PQS analysis. 

The results show that when $\beta/\sqrt{\gamma}$ is small, neither of the methods yield sufficient information, and the best estimate is to assume a vanishing field, and the mean squared error is the corresponding variance of the dog flea model $(0.41\gamma)^2$ as shown in Fig. \ref{beta_figure}

When we increase the probing strength towards $\beta\sim \sqrt{\gamma}$, the mean squared error decreases by a factor of three or four and both analyses yield a significantly better estimate. Further increase of the probe strength has the adverse effect and leads to increased estimation errors. This may be ascribed to the fact that the intra-cavity field becomes stronger and (power) broadens the spin response to detuning changes. In all our calculations, the PQS performs slightly or significantly better than the forward filter approach.

\section{Conclusion}

In this article we have presented a hybrid quantum-classical formalism that permits modelling of the dynamics of a quantum system subject to a classically fluctuating perturbation. The classical fluctuations can be embedded in the quantum formalism by a tensor product of the quantum system Hilbert space with the state space of a Hidden Markov Model. This §leads to a density matrix on block diagonal form which solves a standard Lindblad master equation. Adding observations yields a stochastic master equation which effectively filters the probability distribution of the classical fluctuation. Using quantum measurement theory, we show that forward-backward classical filters for HMMs attain a similar form in quantum sensing based on the past quantum state formalism. We demonstrated use of the formalism for a simple spin tracking the value of a diffusing magnetic field but emphasize the general nature of the method and its ability to deal with more complex noise models and more advanced probe systems \cite{Wang}.  

\begin{acknowledgments}
This work was supported by  the Danish National Research Foundation through the Center of Excellence for Complex Quantum Systems (Grant agreement No. DNRF156) and the  European  QuantERA grant C'MON-QSENS!, by Innovation Fund Denmark Grant No. 9085-00002.
\end{acknowledgments}

\end{document}